\documentclass{article}

\usepackage{epsfig}

\begin{document}

\title{Gamma-spectrometric uranium age-dating using intrinsic efficiency calibration}

\author{%
Cong Tam Nguyen\thanks{E-mail: tam@iki.kfki.hu}
\ and J\'ozsef Zsigrai\thanks{E-mail: zsigrai@sunserv.kfki.hu}\\
{\small \it Institute of Isotopes of the Hungarian Academy of Sciences}\\
{\small \it H-1525 Budapest, P.O. Box 77, Hungary} }

\maketitle

\begin{abstract}
A non-destructive, gamma-spectrometric method for uranium age-dating is presented which is applicable to material of any physical form and geometrical shape. It relies on measuring the daughter/parent activity ratio $^{214}$Bi/$^{234}$U by low-back\-ground, high-resolution gamma-spectrometry using intrinsic efficiency calibration. The method does not require the use of any reference materials nor the use of an efficiency calibrated geometry.



\end{abstract}

\section{Introduction}

If illicit trafficking of nuclear materials is detected, nuclear forensic investigations are carried out to determine the possible origin of the material. One of the parameters to be determined by the methods of nuclear forensic science is the age of the sample, together with the isotopic composition of the material, its physical form, chemical composition etc. In addition, in view of the Fissile Material Cut-Off Treaty knowing the age of isotopically enriched uranium is important for identifying newly produced materials.

The purpose of the present work is to develop a non-destructive, gamma-spectro\-metric method for  age-dating of homogeneous uranium samples, based on using intrinsic efficiency calibration. Other methods for age dating of Uranium samples are mass spectrometry \cite{WalleniusU}, \cite{MorgensternU}, and alpha-spectrometry, which are destructive methods. The gamma-spectrometric method described in \cite{Tam}  and the method given in the present work offer simple non-destructive alternatives.

Gamma-spectrometric age determination of Uranium samples is based on measuring the activity ratio $^{214}$Bi/$^{234}$U by high-resolution gamma-spectro\-metry in low-background, assuming that the daughter nuclides have been completely removed during last separation or purification of the material \cite{Tam}. $^{214}$Bi is a daughter of  $^{234}$U, which decays through $^{230}$Th to $^{226}$Ra, which in turn decays to $^{214}$Bi through three short-lived nuclides. The time needed for secular equilibrium between $^{226}$Ra and $^{214}$Bi to be achieved is about 2 weeks, so it can be assumed that the activities of $^{226}$Ra and $^{214}$Bi are equal at the time of the measurement. Therefore, using the law of radioactive decay the activity ratio $^{214}$Bi/$^{234}$U at time $T$ after purification of the material may be calculated as \cite{Tam}

\begin{eqnarray}
{{A_{Bi214}(T)}\over {A_{U234}(T)}}={{A_{Bi214}(T)}\over {A_{U234}(0)}}= {{A_{Ra226}(T)}\over {A_{U234}(0)}}= 
\lambda_2\lambda_3\biggl[{{e^{-\lambda_1T}}\over {(\lambda_2-\lambda_1)(\lambda_3-\lambda_1)}}+\nonumber\\
{{e^{-\lambda_2T}}\over {(\lambda_1-\lambda_2)(\lambda_3-\lambda_2)}}
+{{e^{-\lambda_3T}}\over {(\lambda_1-\lambda_3)(\lambda_2-\lambda_3)}}\biggr]\ \label{decay},
\end{eqnarray}
where $A_{U234}(0)$ denotes the activity of $^{234}$U at time $T=0$ while $\lambda_1$, $\lambda_2$ and $\lambda_3$ are the decay constants of $^{234}$U, $^{230}$Th and $^{226}$Ra, respectively. Because $\lambda_1, \lambda_2, \lambda_3 <<1$, formula (\ref{decay}) can be developed into a Taylor series around $T=0$, so it can be approximated as
\begin{equation}
{{A_{Bi214}(T)}\over {A_{U234}(T)}} = {1\over 2}\lambda_2\lambda_3 T^2 \ .\label{quadraticage}
\end{equation}
This equation can be used for calculating the age, $T$, of uranium samples after the activity ratio $^{214}$Bi/$^{234}$U has been determined by gamma spectroscopy. 

The activity ratio $^{214}$Bi/$^{234}$U can be determined in several ways from the gamma spectra of the investigated sample. A very precise and reliable way would be to use a reference material of approximately same age, isotopic composition and form as the investigated sample. The samples encountered in illicit trafficking, however, are usually such that an appropriate reference material cannot be found.

Another approach is represented by the method described in \cite{Tam}, that does not require any reference materials but uses the absolute efficiency of the detector determined by ``point-like'' standard sources.  That method is applicable whenever the sample can be approximated to be ``point-like'' or if it has a well defined geometrical shape, for which an efficiency calibrated geometry can be constructed and for which the self-attenuation can also be accounted for.

In the present work a third approach is followed. Here intrinsic efficiency calibration was used to determine the activity ratio $^{214}$Bi/$^{234}$U, without the use of any standard or reference materials. The efficiency calibration is ``intrinsic'' in the sense that a relative efficiency curve is determined from the same spectrum as the measured activity ratios, so the calibration is ``intrinsic'' to the spectrum. The aim of using intrinsic efficiency calibration is to develop a method appropriate for age-dating of homogeneous uranium samples of any arbitrary physical form and shape. The idea is to use the peaks of $^{234m}$Pa, which is a short-lived daughter of $^{238}$U, to construct a relative efficiency curve, and to determine the activity ratio $^{214}$Bi/$^{238}$U. Furthermore, the activity of $^{234}$U is determined relative to $^{235}$U. If the activity ratio $^{235}$U/$^{238}$U is also known one can calculate the activity ratio $^{214}$Bi/$^{234}$U and thus the age of the sample can be obtained. 

In the research described in this work U$_3$O$_8$ powder with a relative isotope abundance in $^{235}$U of 36\% 
was used for the measurements. Gamma spectra of the samples were taken in the 0-300 keV region under standard laboratory conditions and in the 0-2 MeV region in low background. From each spectrum a relative efficiency curve was constructed, using the peaks of $^{235}$U in the first case, and the peaks of $^{234m}$Pa in the second case.

The intensity of the peaks of $^{234m}$Pa can be accurately measured within a reasonable measurement time (1-2 days) if the relative $^{235}$U abundance of the material is less than approximately 90\%, provided a sufficient amount of the investigated material is available (depending on detector efficiency). It should be noted that for lower $^{235}$U abundances the amount of $^{234}$U (and therefore of $^{214}$Bi) is lower as well, so the corresponding activity is more difficult to measure and the uncertainty caused by the variation of the natural background becomes more expressed. In addition, a Compton  background caused by the peaks of $^{238}$U daughters is also present in the spectrum, disturbing the evaluation of the $^{214}$Bi peaks. Therefore there exists a lower limit on the $^{235}$U abundance of the material whose age can be determined by gamma-spectrometry, depending, of course, on the amount and the age of the material, detector efficiency and background level. For example, with our present equipment the age of 10 grams of uranium oxide powder having natural uranium isotopic composition can only be measured if it is more than about 70 years old. On the other hand, it has been estimated that with a fairly average well type detector, if 10 g of uranium oxide is used for the measurement, the lower limit for age dating could be around 30 years for natural uranium, and as low as 2 years for uranium with 90\% of $^{235}$U. A detailed study of the limits for uranium age dating by gamma spectroscopy is the subject of further research.

The age of the material calculated using the described method was compared to the age obtained by using the absolute efficiency of the detector. The results of the two different methods coincide with each other, within the error limits, confirming the reliability of the method presented here.

The structure of this paper is as follows. In section \ref{Experimental} the experimental setup and the procedures for measuring the activity ratios $^{234}$U/$^{235}$U and $^{214}$Bi/$^{238}$U are presented, together with a remark on determining the activity ratio $^{235}$U/$^{238}$U. The results obtained using these procedures are given in section \ref{agesection}, where the age of the investigated material is also calculated.  In section \ref{conclusion} a summary of this work is given with some thoughts about directions for further development of the method.

\section{Experimental}
\label{Experimental}

Four samples of the investigated U$_3$O$_8$ powder, having different thicknesses (and therefore different masses), were prepared with masses of approximately 0.5, 2, 5 and 10 g. The four different-mass samples of the material were each held in a thin, closed polyethylene cylinder with an inner diameter of 2.9 cm. The thickness of the layer of U$_3$O$_8$ powder on the bottom of the polyethylene cylinder varied between 0.04 and 0.92 cm, depending on the mass.
For determining the activity ratio $^{234}$U/$^{235}$U the spectra of the four samples of the assayed material were taken by planar HPGe detector, whereas the activity ratio $^{214}$Bi/$^{238}$U was measured by a coaxial Germanium detector in a low-background chamber.

\subsection{Measuring the activity ratios $^{234}$U/$^{235}$U and $^{235}$U/$^{238}$U}

In order to compare different instruments, the low-energy spectra were taken by two planar high-purity Germanium detectors separately (Canberra GL2020 with active diameter of 50.5 mm, thickness of 20 mm and active surface of 2000 mm$^2$ and Ortec GLP-10180/07 with active diameter of 10 mm and  length of 7 mm). Both detectors were placed in a vertical position, but the ``large'' (50.5 mm diameter) detector was looking upward, while the ``small'' (10 mm diameter) detector was positioned in a downward looking position. 

\begin{figure}[htbp]
\begin{center}
\epsfig{figure=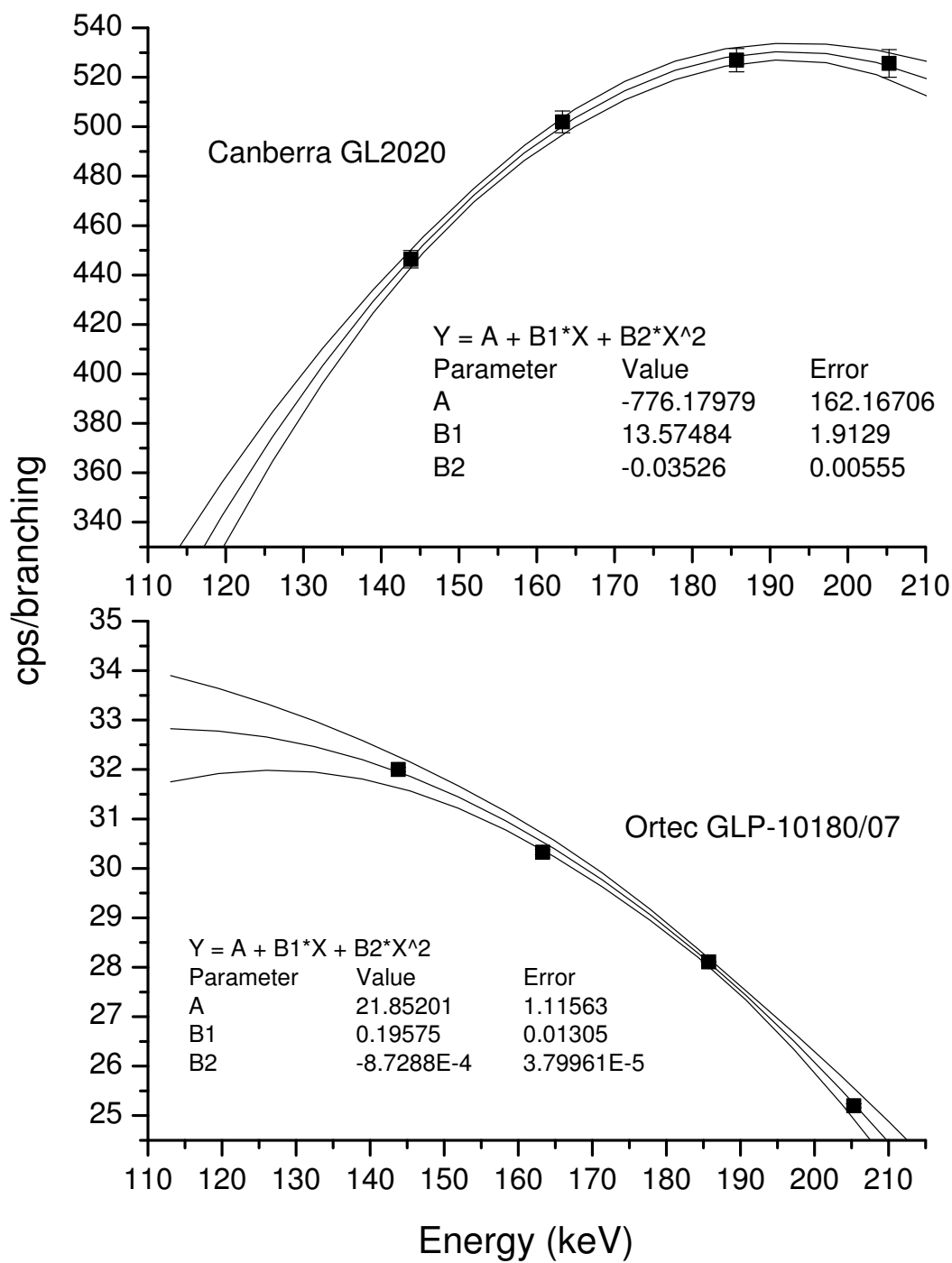, width= 14 cm}
\caption{Relative efficiency curves of the planar HPGe detectors measured using a 10 g sample. The upper graph is for the ``large'', while the lower one is for the ``small'' detector. The 67\%
confidence bands are shown. The points represent the count rates at the given peak energies  of $^{235}$U divided by the corresponding emission probabilities taken from Reference \cite{IAEAdata}. Note that the error bars are not visible on the lower graph because they are smaller than the size of the points}
\label{RelEffPlanar}
\end{center}
\end{figure}

The samples were placed, one after another, at a fixed distance (a few centimeters) from the detector cap and the gamma peak of $^{234}$U at 120.9 keV was observed. The spectrum acquisition lasted until the statistical error of the 120.9 keV line dropped below 2 \%
. The relative efficiency curves of the two detectors are shown on Fig. \ref{RelEffPlanar}.

If the sample is not ``thin''\footnote{Note that the concept of a``thin'' sample depends on the matrix and density of the sample and to some extent even on the detector efficiency and measurement configuration.}, the activity ratio $^{234}$U/$^{235}$U of the material can be reliably determined using the ``U235'' \cite{U235} or the ``MGAU'' \cite{MGAU} code. For thin samples, such as the samples used in this research, however, the results of the U235 and the MGAU codes contain large systematic errors \cite{BerlizovTryshyn}. Therefore, in this research these codes were not used for determining the $^{234}$U content of the material.

In the present work the activity ratio $^{234}$U/$^{235}$U was determined using  relative efficiency calibration, as 

\begin{equation}
{A(^{234}U)\over {A(^{235}U)}}={{I_{120.9}/B_{120.9}}\over{f(120.9keV)}} \label{U234perU235}
\end{equation}
where $A(^{234}U)$ and $A(^{235}U)$ denote the activities of $^{234}$U and $^{235}$U, respectively, $I_{120.9}$ is the intensity of the 120.9 keV line of $^{234}$U, $B_{120.9}$ is its emission probability, while $f(120.9keV)$ is the value of the relative efficiency function of the detector at 120.9 keV. The function $f(E)$ is obtained by fitting a second order polynomial to the relative efficiencies at the 143.8, 163.3, 185.7 and 205.3 keV peaks of $^{235}$U.

The activity ratio $^{235}$U/$^{238}$U was determined from the spectra taken by the small planar detector. It was calculated as the ratio of the relative isotope abundances of $^{235}$U and $^{238}$U determined using the commercial implementation of the ``U235'' \cite{U235} code, which is part of ORTEC's software package ``MGA++'' \cite{MGA++}.
The values we obtained for the $^{235}$U abundance using the U235 code do not depend on the thickness of the sample, confirming the findings of \cite{BerlizovTryshyn}, so we calculated the relative isotope abundance of $^{235}$U and $^{238}$U as the weighted average of the values obtained by measuring the four different-mass samples. They were found to be $36.59\pm	0.01\%$ and $63.08 \pm 0.01\%$, respectively. This corresponds to an $^{235}$U/$^{238}$U activity ratio of $3.664\pm 0.001$ Bq/Bq.

\subsection{Measuring the activity ratio $^{214}$Bi/$^{238}$U}

The activity ratio $^{214}$Bi/$^{238}$U was measured in a low-background iron chamber using a 150 cm$^3$ coaxial germanium detector (``PIGC 3520'' Intrinsic Coaxial Detector manufactured by PGT) having 34.1 \%
relative efficiency (at 1332 keV measured at 25 cm source-detector distance, relative to a 3''$\times$ 3'' NaI(Tl) detector). The wall thickness of the iron chamber is 20 cm and its inner dimensions are $120\times 60\times 120$ cm (height $\times$ width $\times$ length). The detector was standing in vertical position and the samples were placed, one after another, below the detector so that the bottom of the sample holder was at 6 cm from the detector cap. A 4.2 mm lead absorber was attached to the detector cap, in order to reduce the dead-time caused by the high count rate of the low-energy peaks of $^{235}$U. The presence of the lead absorber had no significant influence on the background counts.

\begin{figure}[htbp]
\begin{center}
\epsfig{figure=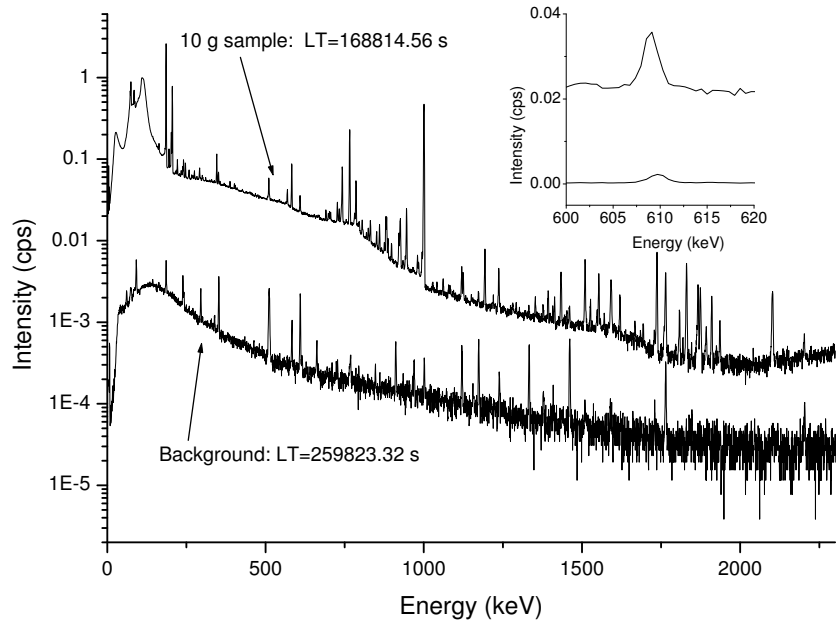, width=11 cm}
\caption{Spectrum of the 10 g sample and of the background, taken by the 150 cm$^3$ coaxial HPGe detector in the low-background chamber, with a 4.2 mm lead absorber in front of the detector cap. The acquisition of the shown spectra lasted for about 2 days for the sample and 3 days for the background.}
\label{CoaxialSpectra}
\end{center}
\end{figure}

\begin{figure}[htbp]
\begin{center}
\epsfig{figure=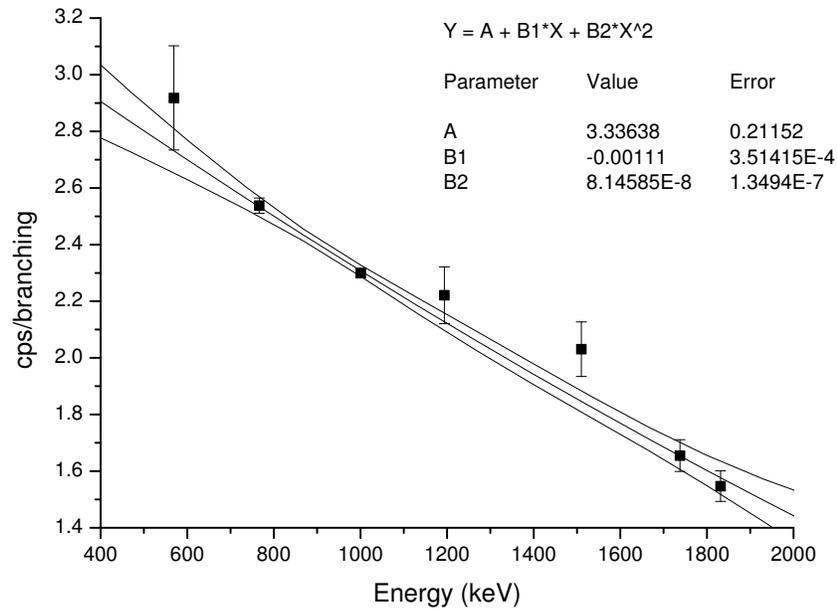, width=11 cm}
\caption{Relative efficiency curve of the coaxial HPGe detector, PGT PIGC 3520, measured using a 10 g sample. (See also the caption below Fig. \ref{RelEffPlanar}.)}
\label{RelEffCoax}
\end{center}
\end{figure}

The gamma-spectra of the samples were taken in the energy range 0-2300 keV (Fig. \ref{CoaxialSpectra}). The small insertion on Fig. \ref{CoaxialSpectra} shows the 609.3 keV peak of $^{214}$Bi. The count rates of the gamma-lines of $^{214}$Bi at 609.3, 1120.3 and 1764.5 keV  were determined from the spectra. Because of small branching ratios and because detector efficiency decreases with energy, the statistical errors for the counts of the last two gamma rays were significant even for long counting times. In addition, these peaks have to be deconvoluted from the peaks of $^{234m}$Pa at 1120.6 keV and 1765.5 keV, which also decreases the reliability of the results obtained for the counts at these energies. Therefore, the 1120.3 and 1764.5 keV  lines were not used in calculating the activity ratio $^{214}$Bi/$^{238}$U.

The activity ratio $^{214}$Bi/$^{238}$U was calculated from the 609.3 keV peak of $^{214}$Bi using the relative efficiency curve obtained from the peaks of $^{234m}$Pa. $^{234m}$Pa is assumed to be in equilibrium with its parent $^{238}$U, so its activity is equal to the activity of $^{238}$U. Analogously to formula (\ref{U234perU235}) one may write
\begin{equation}
{A(^{214}Bi)\over {A(^{238}U)}}={{I_{609.3}/B_{609.3}}\over{F(609.3keV)}} \label{BiperU}
\end{equation}
where $A(^{214}Bi)$ and $A(^{238}U)$ denote the activities of $^{214}$Bi and $^{238}$U, respectively, $I_{609.3}$ is the intensity of the 609.3 keV line of $^{214}$Bi, $B_{609.3}$ is its emission probability, while $F(609.3keV)$ is the value of the relative efficiency function of the detector at 609.3 keV. The function $F(E)$ is obtained by fitting a second order polynomial to the relative efficiencies at the 569.30, 766.37, 1000.99, 1193.69, 1510.20, 1737.73 and 1831.36 keV peaks of $^{234m}$Pa.

It can be clearly seen in Fig. \ref{CoaxialSpectra} that in the 10 g case the count rate of the 609.3 keV peak is much higher than the background, but there is also a high Compton-continuum in the spectrum of the 10 g sample, coming from the higher energy peaks of uranium daughters. The Compton-continuum unfortunately disturbs the evaluation of the relevant peaks. In order to facilitate a more accurate evaluation of the $^{214}$Bi peaks, using a detector having a higher peak-to-Compton ratio would be desirable.

\section{Determining the age of the material}
\label{agesection}

In order to calculate the age of the sample, one needs the activity ratio $^{214}$Bi/$^{234}$U. It is calculated from the results of the above described measurements as

\begin{equation}
{A(^{214}Bi)\over {A(^{234}U)}}={A(^{214}Bi)\over {A(^{238}U)}}\bigg({A(^{235}U)\over {A(^{238}U)}}
{A(^{234}U)\over {A(^{235}U)}}\biggr)^{-1} .
\end{equation}
The age of the material was calculated from equation (\ref{quadraticage}) for each sample separately, as presented in Table \ref{tab:results}. The values $T_{1/2}(^{230}Th)=7.538\times 10^4$ years and $T_{1/2}(^{226}Ra)=1600$ years taken from reference \cite{TORI} for the half life of $^{230}$Th and $^{226}$Ra, respectively, were used in the calculations.

\begin{table}[htbp]
\caption{Results of the measurements obtained using four different-mass samples of the investigated material. The age was calculated from formula (\ref{quadraticage}), using the half lives form reference \cite{TORI}.}
	\begin{center}
		\begin{tabular}{cccccc}
			\hline
			Mass (g)& \multicolumn{3}{c}{Activity ratios}& \multicolumn{2}{c}{Age (years)} \\
			& $^{214}$Bi/$^{238}$U (Bq/Bq)&  \multicolumn{2}{c}{$^{234}$U/$^{238}$U (Bq/Bq)}& & \\
			 & &Small det.& Large det. &Small det.& Large det.\\
			 \hline
			 0.52 & 2(1)$\times 10^{-4}$ & 106(4) &  101(3) & 32(10)& 33(10)\\
			 2.33 & 3.7(6)$\times 10^{-4}$ & 100(2) &  99(9) & 43(4)& 43(6)\\
			 5.55 & 3.6(5)$\times 10^{-4}$ & 97(6) &  99(10) & 43(5)& 43(5)\\
			 10.49 & 3.6(3)$\times 10^{-4}$ & 97(7) &  102(9) & 43(4)& 42(4)\\
			\hline
		\end{tabular}
	\end{center}
	\label{tab:results}
\end{table}

It can be seen from Table \ref{tab:results} that the results obtained using different amounts of the investigated material agree very well with each other, except for the case of the 0.52 g sample. In the case of the smallest sample the error of determining the count rate of $^{214}$Bi is very high, because in this case the relative contribution of the background to the total count rate at 609.3 keV is very high (about 70\%). That is, the error of the count rate at 609.3 keV is dominated by the fluctuations of the background. Accordingly, the results of repeated measurements with the smallest sample placed at 6 cm distance from the detector have shown a big dispersion. The influence of the background can be lowered by placing the sample closer to the detector, so that the relative contribution of the background to the total counts becomes smaller. Therefore, another measurement with the 0.52 g sample was made, but this time placing the sample in such a way that the sample holder was touching the lead shielding. In this measurement the relative contribution of the background to the total counts of the 609.3 kev line was about 50 \%
and the activity ratio $^{214}$Bi/$^{238}$U was calculated to be $4 \pm 1 \times 10^{-4}$ Bq/Bq, resulting in $46\pm 8$ years for the age of the sample. Although this value is close to the value obtained by using larger amounts of the investigated material, its uncertainty is still quite high. Therefore, the measurement of the age of small-amount samples and a detailed study of background fluctuations deserve further attention.

The agreement of the results obtained using different-mass (i.e. different-thickness) samples already indicates that the applied method is not sensitive to the measurement configuration. In order to confirm that the results do not depend on the measurement geometry, the activity ratio $^{214}$Bi/$^{238}$U in the 10 g sample was also determined in two additional, modified measurement geometries. First the sample was placed off-center with respect to the axis of the detector, at approximately 2 cm from the detector axis. Then the cylindrical sample holder was positioned again to the detector axis, but in such a way, that the axis of the cylinder was at an angle of approximately 45 degrees with respect to the axis of the detector. The activity ratio $^{214}$Bi/$^{238}$U was found to be  $(3.6 \pm 0.3)\times 10^{-4}$ Bq/Bq in the first case and $(3.5 \pm 0.3)\times 10^{-4}$ Bq/Bq in the second case. Both these values agree well with the values listed in Table \ref{tab:results}, confirming the independence of the results on the measurement geometry and on the shape of the sample.

As for the activity ratio $^{234}$U/$^{238}$U, it can be seen that the values obtained by using different detectors are consistent with each other. If the larger detector is used, the counting times needed to achieve the desired counting statistics are consequently shorter, so this detector may be the preferred choice when only a small amount of the investigated material, producing a low count rate, is available. On the other hand, if time is not critical or if a larger amount of the material to be measured is at disposal, the smaller detector might be a better choice. Namely, the smaller detector has better energy resolution, making possible the automatized calculation of the $^{235}$U and $^{238}$U abundance  using the computer codes U235 or MGAU, which require a good energy resolution to produce correct results. Note that the uncertainty associated with the evaluation of the relative efficiency function at 120.9 keV is relatively high, producing an uncertainty of about 9-10 \%
in the value of the activity ratio $^{234}$U/$^{238}$U (see Table \ref{tab:results}), which is much higher than the dispersion of the measured values. Presently further studies are being carried out in order to understand better and reduce the uncertainties of determining the activity ratio $^{234}$U/$^{238}$U.

For comparison, the age of the investigated material was also calculated by using the absolute efficiency of the detector, as in reference \cite{Tam}. Denoting the absolute efficiency of the detector at energy $E$ by $\epsilon_E$, the activity, $A$, of a given isotope can be calculated as
\begin{equation}
A={I_E \over{B_E\epsilon_E}}\ , \label{abs-activity}
\end{equation}
where  $B_E$ and $I_E$ are, respectively, the emission probability and measured count rate of the observed gamma line of energy $E$, corresponding to the given isotope.
In order to account for the self absorption of the 120.9 keV line within the sample, the procedure described in \cite{Tam} was applied. The count rate corrected for self-absorption is given in Table \ref{tab:absolute}. In the case of the 609 keV line of  $^{214}$Bi the uncertainty caused by self-absorption was negligible compared to the statistical error of the observed peaks. Namely, the layer of U$_3$O$_8$ powder used in the measurements is practically transparent for the gamma photons at these energies and there was no need to correct for self absorption. This was also confirmed by measurements, i.e., the count rates per unit mass for the different-mass samples were approximately the same. A much more important source of error is the random fluctuation of the background level. In the case of the 10 g sample the count rate is the highest, so in this case the relative uncertainty produced by the fluctuation of the background level is the lowest. Therefore, only the results of the measurements with the 10 g sample were used for calculating the activity of $^{214}$Bi (Table \ref{tab:absolute}). The age obtained using the activities given in Table \ref{tab:absolute} is $45\pm 4$ years, which, within the error limits, coincides with the value obtained using intrinsic efficiency calibration. 

\begin{table}[htbp]
\caption{The activities of $^{214}$Bi and $^{234}$U obtained by using the absolute detector efficiency. The emmission probabilities were taken from references \cite{IAEAdata} and \cite{TORI} for  $^{234}$U and $^{214}$Bi, respectively.}
	\begin{center}
		\begin{tabular}{cccccc}
			\hline
			{\bf Nuclide} & \parbox{43pt}{\center{{\bf Energy}\\ (keV})} & \parbox{65pt}{\center{{\bf Emmission probability}\\ (\%)}}&\parbox{50pt}{\center{{\bf Detector\\ efficiency}\\ (\%)}} & \parbox{60pt}{\center{\bf Count rate per unit mass\\ (cps/g)}}&  \parbox{50pt}{\center{{\bf Activity}\\ (Bq/g)}}\\
			\hline
			$^{234}$U & 120.90 & 0.0342(5) & 0.730(15)& 1.46(3) &$5.9(3)\times 10^4$\\
			\hline
			$^{214}$Bi &  609.3 & 46.1(5) &  0.39(1) & $4.2(3)\times 10^{-3}$& 2.3(2)\\
			\hline
		\end{tabular}
	\end{center}
	\label{tab:absolute}
\end{table}

\section{Conclusion}
\label{conclusion}

In this work a gamma-spectrometric method for determining the age of isotopically enriched uranium samples with medium $^{235}$U abundance is presented. The method does not require the use of standard samples of known age, nor the knowledge of the absolute detector efficiency and it is applicable for determining the age of uranium samples of any arbitrary geometrical shape. The age of the investigated sample is calculated from the activity ratio $^{214}$Bi/$^{234}$U, using formula (\ref{quadraticage}). In the present work intrinsic efficiency calibration was used to determine the activity ratios $^{234}$U/$^{235}$U and $^{214}$Bi/$^{238}$U, from which the activity ratio $^{214}$Bi/$^{234}$U can also be calculated, provided that the activity ratio $^{235}$U/$^{238}$U is known. The activity ratio $^{235}$U/$^{238}$U can be determined from the same spectrum as the activity ratio $^{234}$U/$^{235}$U, using standard methods based on intrinsic efficiency calibration, such as, e.g., ``U235'' and ``MGAU''.

The results obtained using the described method were confirmed by a method which uses the absolute efficiency of the detector. It can be concluded that the method described in this work is a reliable alternative for determining the age of isotopically enriched uranium samples with a medium $^{235}$U abundance, having any arbitrary shape.

Presently the background fluctuations are being studied in detail, in order to find a way to reduce them and thus improve the precision of evaluating the 609.3 keV peak of $^{214}$Bi. Another important possibility for extending the capabilities of the method would be the use of a detector with a higher efficiency and thus higher peak-to-Compton ratio for the low-background measurements. Finally, the uncertainties associated with determining the activity ratio $^{234}$U/$^{235}$U by intrinsic efficiency calibration also deserve further attention.

\section*{Acknowledgment}

This research has been partially supported by the Hungarian Atomic Energy Authority under the research and development contract OAH/\'ANI-ABA-02/04. The authors wish to thank Laszlo Lakosi for checking the manuscript.

\end{document}